\documentclass[%
 aps,prb,reprint,
superscriptaddress,
groupaddress,
floatfix,
]{revtex4-1}
\bibpunct{[}{]}{,}{n}{}{}
\usepackage{graphicx}
\usepackage{dcolumn}
\usepackage{bm}

\usepackage{color}
\usepackage[none]{hyphenat}
\usepackage{amsmath} 
\usepackage{amssymb} 

\usepackage{xcolor}
\usepackage[normalem]{ulem}

\begin{document}
\newcommand{\bsts}{Bi$_{1.5}$Sb$_{0.5}$Te$_{1.7}$Se$_{1.3}$ }
\newcommand*\tred[1]{\textcolor{red}{#1}}
\newcommand{\Zsym}{z $\rightarrow$ -z }

\preprint{APS/123-QED}

\title{Proximity-Induced Spin-Orbit Coupling in \\Graphene-\bsts Heterostructures}

\author{S. Jafarpisheh}
\affiliation{JARA-FIT and 2nd Institute of Physics, RWTH Aachen University, 52074 Aachen, Germany}
\affiliation{Peter Gr\"unberg Institute (PGI-9), Forschungszentrum J\"ulich, 52425 J\"ulich, Germany}
\author{A.W. Cummings}
\affiliation{Catalan Institute of Nanoscience and Nanotechnology (ICN2), CSIC and BIST, Campus UAB, Bellaterra, 08193 Barcelona, Spain}
\author{K. Watanabe}
\affiliation{National Institute for Materials Science, 1-1 Namiki, Tsukuba, 305-0044, Japan}
\author{T. Taniguchi}
\affiliation{National Institute for Materials Science, 1-1 Namiki, Tsukuba, 305-0044, Japan}
\author{B. Beschoten}
\thanks{ e-mail: bernd.beschoten@physik.rwth-aachen.de}
\affiliation{JARA-FIT and 2nd Institute of Physics, RWTH Aachen University, 52074 Aachen, Germany}
\author{C. Stampfer}
\affiliation{JARA-FIT and 2nd Institute of Physics, RWTH Aachen University, 52074 Aachen, Germany}
\affiliation{Peter Gr\"unberg Institute (PGI-9), Forschungszentrum J\"ulich, 52425 J\"ulich, Germany}

\date{\today}

\begin{abstract}

The weak intrinsic spin-orbit coupling in graphene can be greatly enhanced by proximity coupling.
Here we report on the proximity-induced spin-orbit coupling in graphene transferred by hexagonal boron nitride (hBN) onto the topological insulator \bsts (BSTS) which was grown on a hBN substrate by vapor solid synthesis. Phase coherent transport measurements, revealing weak localization,
 allow us to extract the carrier density-dependent phase coherence length $l_\phi$. While $l_\phi$ increases with increasing carrier density in the hBN/graphene/hBN reference sample, it decreases in graphene/BSTS due to the proximity-coupling of BSTS to graphene. The latter behavior results from D'yakonov-Perel-type spin scattering in graphene with a large proximity-induced spin-orbit coupling strength of at least 2.5~meV.
\end{abstract}

\pacs{}
\maketitle


Graphene (Gr) has become a promising material for spintronics  due to its long spin lifetimes and spin diffusion lengths~\cite{Han2014Oct,Roche2DMat2015,DroegelerNano2014,Drogeler2016Jun,GuimaraesPRL2014,AynesNano2016,Drogeler2017Oct}. Tailoring the spin-orbit coupling (SOC), a key ingredient for spin manipulation, can bring Gr one step closer to its integration into functional devices. Various experimental methods such as hydrogenation~\cite{Kaverzin2015Apr}, fluorination~\cite{Avsar2DMat2015} and heavy adatom adsorption~\cite{JiaPRB2015} have been proposed. However, as a major drawback, these methods often deteriorate the transport properties of Gr. Another approach is the use of two-dimensional materials such as transition metal dichalcogenides (TMDs) which exhibit large intrinsic SOC~\cite{Garcia2018May,JiangNano2015,AvsarNature2014,WangNature2015,OmarPRB2018,WangPRX2016}. These materials not only allow for high carrier mobilities in Gr~\cite{Banszerus2017Feb}, but also induce SOC into Gr by the interface proximity effect. Indeed, transport measurements on Gr proximity-coupled to TMDs have shown an enhanced SOC in Gr by several orders of magnitude, with the potential to allow for new device functionalities~\cite{AvsarNature2014,WangNature2015,WangPRX2016,OmarPRB2018,Yang2DMat2016,Yan2016Nov,Benntez2017Dec,Dankert2017Jul,Zihlmann2018Feb,Banszerus2017Feb}.
Interesting alternative materials are topological insulators (TI), which offer a unique electronic band structure with conducting surface states where electron spins are locked to their momentum~\cite{Ando2013Sep,Vaklinova2016Apr}. Recently, there have been several theoretical studies predicting TI--to--Gr hybridization and transfer of the TI spin texture to Gr~\cite{Cao2DMat2016,JinPRB2013,VegaPRB2017,ZhangPRL2014,PopovPRB2014,Song2018Mar}. The interface of the two materials has been studied by angular-resolved photoemission~\cite{LeeACSNano2015} as well as vertical transport measurements~\cite{BeulePRB2017,ZalicPRB2017}. In addition, anomalous quantum transport properties of Gr/Bi$_2$Se$_3$ suggests strong electronic coupling between the two materials~\cite{ZhangACSNANO2017}. However, phase coherence transport in TI/Gr hybrid systems remains unstudied.

Here we report on weak localization (WL) studies of heterostructures based on Gr and \bsts(BSTS) encapsulated in hexagonal boron nitride (hBN). Comparing the 
 carrier density dependence of the extracted phase coherence length with that of Gr encapsulated in hBN gives insight into the SOC induced in Gr by BSTS. While the phase coherence length for hBN/Gr/hBN (Gr/hBN) increases with increasing charge carrier density, it strongly
  decreases for hBN/Gr/BSTS/hBN (Gr/BSTS). This decrease indicates the dominance of D'yakonov--Perel (DP) spin relaxation as a result of proximity--induced SOC. We estimate a lower limit of 2.5 meV for the strength of the proximity--induced SOC in the Gr/BSTS heterostructure.

\begin{figure*}[t]%
	\includegraphics*[width=.9\textwidth]{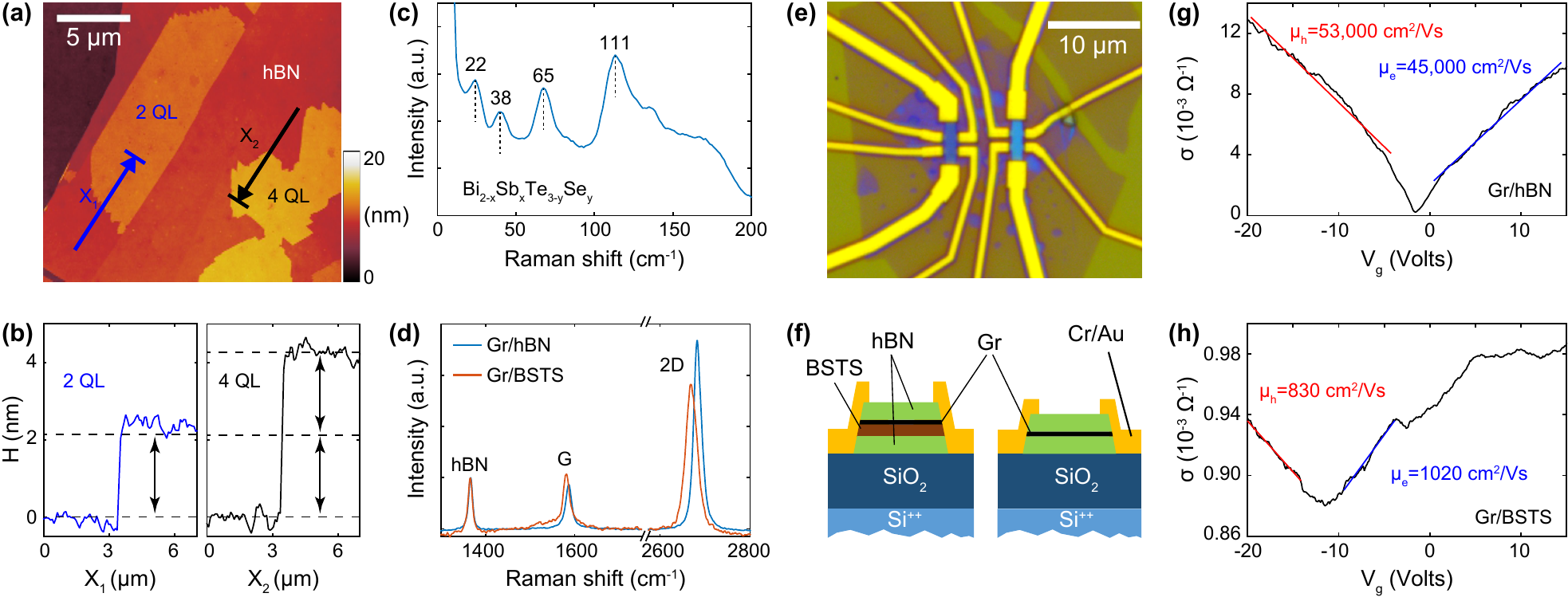}
	\caption{\label{fig:figure1}
		(a) SFM image of BSTS grown on hBN. (b) Height profiles at the edges of the flakes extracted from SFM scan. (c) Raman spectra of a typical BSTS layer deposited on hBN.
		(d) Raman spectra of the Gr flake used for transport measurements on BSTS (red) and on hBN (blue). (e) Optical image of the two devices (Gr/BSTS and Gr/hBN) used for transport measurements with Cr/Au contacts. (f) Schematic view of both heterostructures. (g) and (h) Conductivity of Gr/hBN and Gr/BSTS devices as a function of gate voltage $V_g$ with $\mu_e$ and $\mu_h$ being the respective electron and hole mobilities.}
\end{figure*}

BSTS layers were deposited on exfoliated hBN flakes resting on SiO$_2$/Si$^{++}$
using a catalyst--free vapor--solid synthesis from \bsts crystals following Ref.~[\citenum{Ockelmann2015Aug}] (see Supplementary Materials~\cite{SupMat}). 
 We grow BSTS crystals with a thickness of only a few quintuple layers (QLs) to minimize parasitic charge transport channels through the BSTS layer in the Gr/BSTS devices. Fig.~1(a) shows a scanning force microscope (SFM) image of typical BSTS crystals grown on hBN showing step--less surfaces (Fig.~1(b)) confirming a homogeneous layer--by--layer growth. Raman spectra of BSTS flakes (Fig.~1(c)) show three active modes with frequencies lower than 100 cm$^{-1}$ (2 E and 1 A$_1$ mode), which confirms the formation of BSTS~\cite{Watanabe2014Raman}.
In a second step we exfoliate Gr from natural graphite onto a second SiO$_2$/Si substrate which gets dry-transferred \cite{WangScience2013,BanszeruseSienceA2015,Drogeler2016Jun} on top of the BSTS(2 QLs)/hBN stack to assemble the hBN/Gr/BSTS/hBN heterostructure. The air exposure time of the BSTS 
 prior to the transfer of Gr was limited to a few minutes which minimizes oxidation of its surface layer. This is crucial to allow proximity coupling across the BSTS--to--Gr interface. As the bottom hBN was not completely covered by BSTS  (see e.g.~Fig.~1(a)), parts of the final heterostructure are BSTS--free, resulting in a hBN/Gr/hBN sandwich assembled during the same fabrication step which we use as a reference device.

Raman spectroscopy was used to characterize the Gr flake in both the Gr/hBN and Gr/BSTS regions (Fig.~1(d)). For the latter, the G and 2D peak frequencies ($\omega$) show a red shift ($\Delta\omega_\mathrm{G}\approx 6$ cm$^{-1}$, $\Delta\omega_{\mathrm{2D}}\approx 15$ cm$^{-1}$) which is due to the strain introduced by the BSTS substrate~\cite{LeeNature2012}. The broadening of the 2D peak ($\Delta\Gamma_{\mathrm{2D}}\approx11$ cm$^{-1}$ with $\Gamma$ being the full-width at half-maximum of the peak) can be associated with  higher nm-scale strain variations in the Gr on BSTS compared to hBN~\cite{KoSR2013,NeumannNature2015}. Electrical contacts were fabricated using electron beam lithography followed by metallization with Cr(5~nm)/Au(120~nm) and lift-off (see also Ref.~[\citenum{SupMat}]). An optical image and a schematic cross sectional view of both devices are shown in Figs.~1(e) and 1(f). Transport measurements were performed at a temperature of $T=10$~mK using low-frequency lock-in techniques with a constant current of $1~\mu\mathrm{A}$.

In Figs.~1(g) and 1(h) we show the conductivity $\sigma$ as function of gate voltage $V_\mathrm{g}$ (applied to the Si$^{++}$ layer) for both devices.
From the Drude formula $\sigma=e n \mu + \sigma_0$ where $\sigma_0$ accounts for the parallel conduction channel through the BSTS layer in the Gr/BSTS device, we extract the respective mobilities $\mu$ (numbers are given in both panels) with $e$ being the elementary charge and $n$ the charge carrier density in Gr calculated using the gate lever arm $\alpha$, which is extracted from (quantum) Hall measurements (see below). The drastic difference between the two devices also becomes apparent in their Landau-fan diagrams in Figs.~2(a) and 2(b). The dashed lines follow the Landau levels (LLs) given by  $n = \alpha(V_\mathrm{g}-V_\mathrm{g}^\mathrm{cnp})$, where  
 $V_\mathrm{g}^\mathrm{cnp}$ is the gate voltage of the charge neutrality point (CNP). For Gr/hBN we extract
 $\alpha = 7$~$\times$~$10^{10}$ cm$^{-2}$V$^{-1}$ from
 Hall effect measurements, which fits well with the Landau-fan diagram in Fig.~2(a).
\begin{figure*}[t]%
	\includegraphics*[width=.95\textwidth]{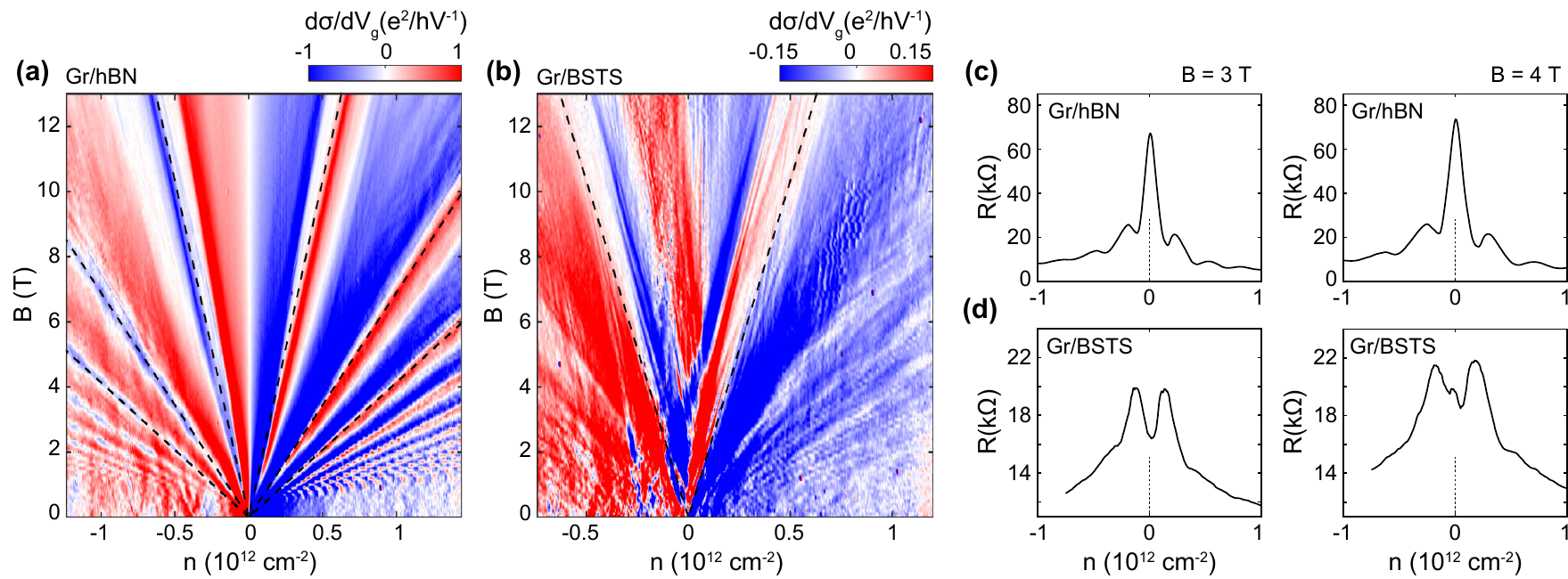}
	\caption{\label{fig:figure2}
		(a) Transconductivity of Gr/hBN vs $n$ and perpendicular $B$-field. The dashed lines indicate filling factors of $\nu=\pm2,$$\pm6,$$\pm10$. (b) Landau fan diagram of Gr/BSTS device. The dashed lines show the filling factor of $\nu=\pm2$. (c-d) Line traces of the corresponding resistivities at $B=3$ and 4~T for (c) Gr/hBN and (d) Gr/BSTS.
	}
\end{figure*}
By comparing to a second reference device with only a BSTS flake (2 QLs) sandwiched by hBN, which did not show any $B$-field-dependent signatures of Landau quantization (see Fig.~S1 in Supplementary Materials~\cite{SupMat}), we conclude that the Landau-fan shown in Fig.~2(b) originates from Gr only. The slopes of the dashed lines allow to extract a gate lever arm of 5~$\times$~$10^{10}$ cm$^{-2}$V$^{-1}$. This smaller value compared to the Gr/hBN device most likely results from screening effects of the BSTS layer which is located between Gr and the gate (see Fig.~1(f)).

The first indication of proximity coupling of BSTS to Gr becomes apparent when comparing the density dependent resistances of both devices for $B$-fields of 3 and 4~T, as shown in Figs.~2(c) and 2(d). While the Gr/hBN device shows the expected peak in the resistance at the CNP, i.e. at $n=0$ (see Fig.~2(c)), which corresponds to the zeroth LL, there is a minimum resistance near the CNP in the Gr/BSTS device up to a B-field of 4~T (see Fig.~2(d)). This indicates a strong modification of the electronic structure of Gr in proximity to BSTS. This unusual behavior has recently been observed in Gr/Bi$_2$Se$_3$ by Zhang et al.~\cite{ZhangACSNANO2017} for negative magnetic fields only. They attributed the strong asymmetry of the magneto-resistance for both positive and negative $B$-fields to the spin texture of the Bi$_2$Se$_3$ surface states which proximity-couple to the Gr states.
We note that we do not observe this asymmetry in our devices~\cite{SupMat}. This is most likely related to the ultra--thin BSTS layer of only 2 QLs, which is much thinner than the threshold thicknesses reported for having decoupled surface states~\cite{LinderPRB2009,ZhangNature2010,TuNature2016}. Nevertheless, the existence of the minimum resistance near the CNP shows BSTS--induced proximity coupling in our devices.

We next 
 discuss how this proximity coupling affects phase-coherent transport. In Fig.~3, we show representative low-field magneto-conductivity data of Gr/hBN (blue curves) and Gr/BSTS (red curves) at both low ($n = 5.5 \times 10^{10}$ cm$^{-2}$) (Fig.~3(a)) and high densities ($n = 1.1 \times 10^{12}$ cm$^{-2}$) (Fig.~3(b)). The increase of conductivity away from $B=0$ is a hallmark of WL, which has been extensively studied in Gr~\cite{MorpurgoPRL2006WL,McCannPRL2006,MorozovPRL2006WL,McCannPRL2012WL,EngelsPRL2014}. While close to the CNP both curves look quite similar (Fig.~3(a)) they become  distinctly different at large densities (Fig.~3(b)). In the following, we analyze our data with the theoretical model proposed by McCann and co-workers~\cite{McCannPRL2006},
\begin{eqnarray}
	\Delta \sigma \left(B\right)&=&\frac{{e}^{2}}{\pi h}\biggl[F\left(\frac{B}{{B}_{\mathrm{\phi} }}\right)-F\left(\frac{B}{{B}_{\mathrm{\phi}}+2{B}_\mathrm{i}}\right)\nonumber\\
	& &-2F\left(\frac{B}{{B}_{\mathrm{\phi} }+{B}_\mathrm{i}+{B}_\mathrm{\ast }}\right)\biggl],
\end{eqnarray}
where $F\left(z\right)=\mathrm{ln}z+\psi \left(\frac{1}{2}+\frac{1}{z}\right)$ and ${B}_{\mathrm{\phi,i,*} }=\frac{\hslash}{4e}{l }_{\mathrm{\phi,i,*}}^{-2}$. Here, $\psi$ is the digamma function and $\l_{\mathrm{\phi}}$, $\l_\mathrm{i}$, $\l_{*}$ are the phase coherence, inter-valley and intra-valley scattering length scales respectively. This model requires three fitting parameters in addition to a pre-factor for adjusting the magnitude of the WL signal. In most measurements the WL signal is superimposed on universal conductance fluctuations (see e.g. blue curve in Fig.~3(b)). As a result, we find a huge uncertainty in the extracted fitting parameters, specifically for $l_\mathrm{i}$ and $l_*$. We therefore restrict the fit to the lowest $B$-field region ($\pm$10 to 15~mT in Gr/BSTS and $\pm$3 to 10~mT in Gr/hBN) and analyze the width and magnitude of the WL signal, which directly determines $l_\phi$. With this approach, we can extract values of $l_\phi$ with decent accuracy (see error bars in Fig.~4). In Fig.~3 we added the respective fitting curves (see dashed lines), showing good agreement at low $B$-fields while deviating from the measurements at higher fields. When including $l_\mathrm{i}$ and $l_*$ into the fitting procedure, the results are in better agreement at higher fields, and they have almost no effect on the values of $l_\phi$. 
We therefore restrict the following discussion to the extracted values of $l_{\phi}$ only.

\begin{figure}[b]%
	\includegraphics*[width=\linewidth]{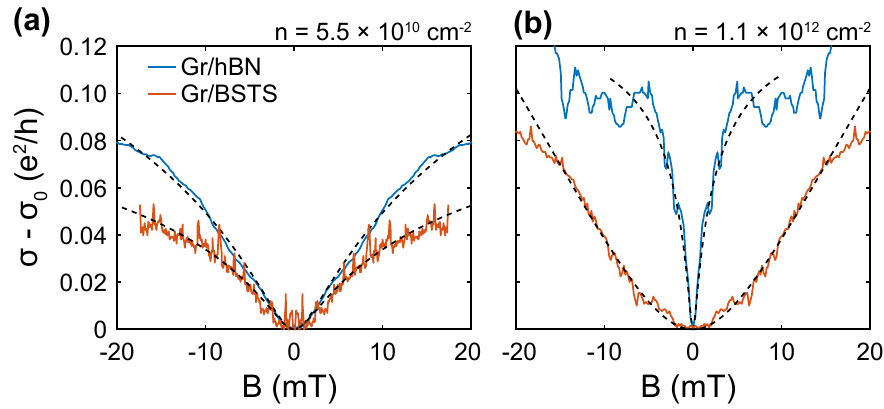}
	\caption{\label{fig:figure3}
			WL measurements Gr/hNB (blue) and Gr/BSTS (red) (a) close to CNP and (b) at high densities ($n = 1.1 \times 10^{12}$ cm$^{-2}$). The dashed lines are fits to the McCann model for WL in Gr for both panels.}
\end{figure}

\begin{figure}[t]%
	\includegraphics*[width=\linewidth]{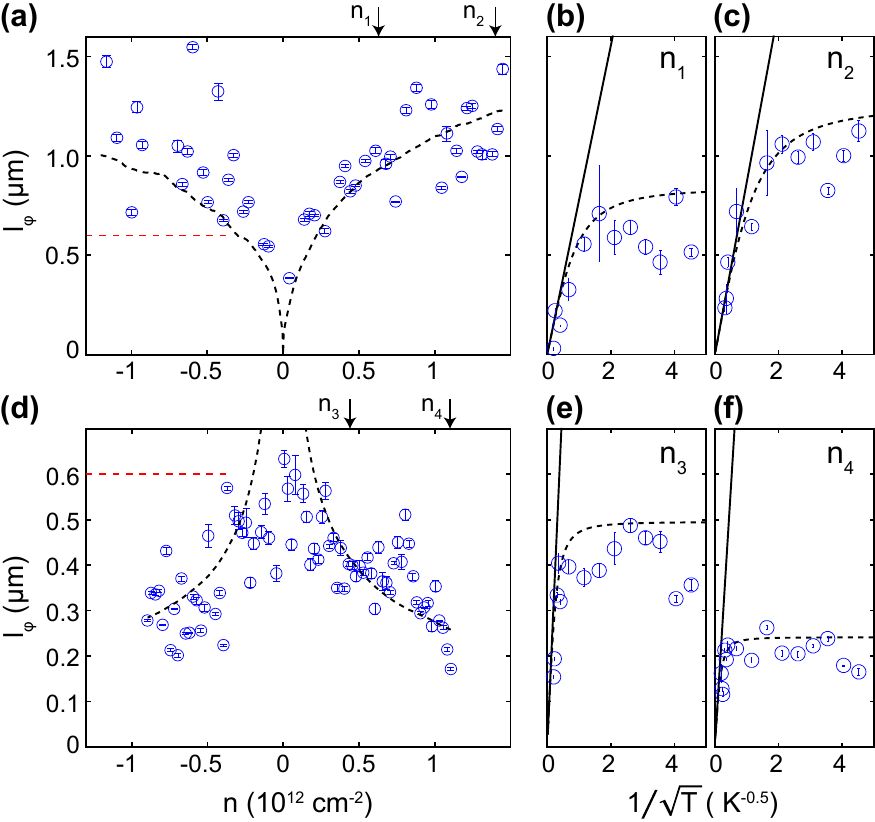}
	\caption{\label{fig:figure4}
		(a) $\l_{\phi}$ vs $n$ taken at 10~mK for Gr/hBN and the modified AAK model (black dashed line). (b) and (c) show $T$ dependence of $\l_{\phi}$ and the AAK model (solid lines) in addition to the modified AAK model (dashed lines) at (b) $n_1 = 6.3 \times 10^{11}$ cm$^{-2}$ and (c) $n_2 = 1.4 \times 10^{12}$ cm$^{-2}$. (d) $\l_{\phi}$ vs $n$ for Gr/BSTS. (e) and (f): $T$ dependence of $\l_{\phi}$ at (e) $n_3 = 4.4 \times 10^{11}$ cm$^{-2}$ and (f) $n_4 = 1.1 \times 10^{12}$ cm$^{-2}$.}
\end{figure}

Figure 4 summarizes the dependence of $l_\phi$ on $n$ and $T$ for both the Gr/hBN (Figs.~4(a) to 4(c)) and the Gr/BSTS (Figs.~4(d) to 4(f)) devices~\footnote{Since universal conductance fluctuations superimpose the WL curves, the extracted values of the phase coherence length are heavily scattered}. The former exhibit the typical increase of $l_\phi$ away from the CNP for both electron ($n>0$) and hole ($n<0$) doping as previously reported~\cite{GeFOP2017,Ki2008Sep}. This behavior is in qualitative agreement with a scattering mechanism based
on electron-electron interactions as predicted by Altshuler-Aronov-Khmelnitsky (AAK),
\begin{eqnarray}
l_{\phi}={\sqrt{D\tau_\phi}} \;\;\; \text{with} \;\;\; \tau_{\phi}=\hslash g_{\square}\left(k_BT\mathrm{ ln}g_{\square} \right)^{-1},
\end{eqnarray}
where $k_\mathrm{B}$ is the Boltzmann constant, $g_\square=\sigma h/e^2$ is the normalized conductivity and $D=v_F g_{\square}/(4\sqrt{n\pi})$ is the diffusion constant with $v_\mathrm{F}$ being the Fermi velocity. However, the extracted values from WL measurements at 10 mK are much smaller than the predictions by AKK.  
The temperature dependence of $l_\phi$ from 25~K down to 10~mK (Figs.~4(b-c)) shows that $l_{\phi}$ is inversely proportional to the square root of $T$ ($l_{\phi}\propto 1/\sqrt{T}$) above 1~K, but saturates at lower temperatures. This saturation has been attributed to 
spin scattering at residual magnetic impurities and their resulting effective local magnetic moments~\cite{AvilaPRL2011,FedorovPRL2013,KochanPRL2014,LundebergPRL2013}.
Following Ref.~[\citenum{EngelsPRL2014}], we therefore include an additional spin scattering 
leading to $\tau_{\phi}^{-1}={\tau_\textbf{s}}^{-1} + k_BT\mathrm{ ln}g_{\square}/{\hslash g_{\square} }$,
where $\tau_\mathrm{s}$ is the spin lifetime. From the $n$ dependent changes of $l_{\phi}\propto \sqrt{\tau_{\phi}}$ in Figs.~4(a) and 4(d) we can now identify the dominant spin scattering mechanisms. The increase of $l_{\phi}$ with increasing $|n|$ for Gr/hBN in Fig.~4(a) can be attributed to spin-flip scattering given by $\tau_\mathrm{s} = \tau_\mathrm{sf}= \beta\left|n\right|$ with $\beta = 7\times 10^{-23}$ cm$^2$s.
As shown by the dashed lines in Figs.~4(a) to 4(c), this assumption gives a good quantitative agreement with all data without any additional adjustable parameters. We extract $\tau_\mathrm{sf}=44$~ps and 100~ps for $n_1 = 6.3 \times 10^{11}$ cm$^{-2}$ ($V_\mathrm{g}=11$~V) and $n_2 = 1.4 \times 10^{12}$ cm$^{-2}$ ($V_\mathrm{g}=20$~V), respectively. These values are consistent with previous reports for Gr~\cite{AvilaPRL2011}.

We now focus on the Gr/BSTS device, which shows a distinctly different $n$ dependence of $l_{\phi}$ in Fig.~4(d). Close to the CNP ($n=0$) $l_{\phi}$ exhibits similar values as the Gr/hBN device (see red lines in Figs.~4(a) and 4(d)). The strong decrease of $l_{\phi}$ with increasing $|n|$ indicates the dominance of a different spin scattering mechanism in the Gr layer, leading to DP-type spin relaxation. As shown in Figs.~4(e) and 4(f), spin scattering also limits phase coherent transport at low $T$ as $l_{\phi}$ becomes $T$  independent.

A comprehensive model for WL and weak antilocalization (WAL) in Gr in the presence of SOC is provided by McCann and Fal'ko~\cite{McCannPRL2012WL}. They consider SOC terms which are symmetric or asymmetric upon $z$/-$z$ inversion.
In symmetric systems spin-orbit scattering is governed by intrinsic and valley-Zeeman SOC while in asymmetric systems, SOC result from Rashba and pseudospin-inversion asymmetries~\cite{Gmitra2016Apr,Kochan2017Apr,Zihlmann2018Feb}.
However, determining both the symmetric and asymmetric contributions from this model requires seven fitting parameters~\cite{McCannPRL2012WL}.
Following the above discussion, our measurements do not allow to extract all of them with reasonable accuracy. Nevertheless, we show in the Supplementary Materials that we can reproduce the WL curve within a larger $B$-field range at certain densities with a rough estimation of each parameter. Based on this analysis, we find a negligible contribution of asymmetric SOC, which is consistent with the absence of WAL at most carrier densities \cite{SupMat}. The remaining symmetric contributions ($\tau_\mathrm{sym}$) can be quantified by studying the saturation behavior and $n$ dependency of $\tau_{\phi}$ at low $T$ ($\tau_{\phi}(T\rightarrow0)\rightarrow\tau_\mathrm{sym}$)~\cite{McCannPRL2012WL}.
Thus, we therefore approximate the dominating spin scattering time by
\begin{eqnarray}
\tau_\mathrm{s}=\tau_\mathrm{sym}=\frac{\hslash^2}{2\lambda_\mathrm{sym}^2}{\tau_\mathrm{p}}^{-1} \;\;\; \text{with} \;\;\; \tau_\mathrm{p}=\frac{\mu h}{2ev_\mathrm{F}\sqrt{\pi}}\sqrt{n},
\end{eqnarray}
where $\tau_\mathrm{p}$ is the momentum scattering time, and $\lambda_\mathrm{sym}$ is the strength of the proximity-induced symmetric SOC.
Fitting results are included in Figs.~4(d) to 4(f) as black dashed lines with $\tau_\mathrm{sym}=4.5$~ps and 2.9~ps for $n_3 = 4.4 \times 10^{11}$ cm$^{-2}$ ($V_\mathrm{g}=11$~V) and $n_4 = 1.1 \times 10^{12}$ cm$^{-2}$ ($V_\mathrm{g}=20$~V), respectively. Compared to the Gr/hBN reference sample, $\tau_\mathrm{s}$ is significantly reduced by the large proximity-induced SOC from BSTS to Gr. With the extracted mobility of $\mu_e \approx 1000 $~cm$^2/\mathrm{(Vs)}$ for Gr in Gr/BSTS (see Fig.~1(h)), we estimate the lower bound of the symmetric SOC strength to be $\lambda_\mathrm{sym} = 2.5$ meV.

The above analysis indicates that spin relaxation in the Gr/BSTS system is dominated by symmetric SOC, which is typically associated with intrinsic SOC. In Gr, intrinsic SOC leads to Elliott-Yafet spin relaxation, such that $\tau_\mathrm{s} \propto \tau_\mathrm{p}$ \cite{Ochoa2012May}. This scaling behavior is at odds with Eq.~3 and Fig.~4, suggesting that intrinsic SOC is not dominant in our Gr/BSTS devices. However, recent work has shown that other forms of SOC can play a role in Gr/TI heterostructures \cite{Song2018Mar}. Depending on the symmetry of the Gr/TI interface, the Gr spin texture can be dominated by valley-Zeeman or by a Rashba-like SOC arising from strong in-plane electric fields, and both of these remain symmetric under $z$/-$z$ inversion. A Valley-Zeeman SOC leads to DP-like spin relaxation $\tau_\mathrm{s} \propto \tau_\mathrm{iv}^{-1}$ with $\tau_\mathrm{iv}$ being the intervalley scattering time \cite{Cummings2017Nov}, while the in-plane Rashba fields lead to typical DP behavior \cite{Song2018Mar}, $\tau_\mathrm{s} \propto \tau_\mathrm{p}^{-1}$. Either or both of these mechanisms could therefore be playing a role in our devices.

In conclusion, phase coherent transport measurements in Gr/BSTS unveil the proximity-induced SOC from BSTS onto Gr. The overall absence of WAL indicates the dominance of SOC terms which are symmetric upon $z$/-$z$ inversion. The decrease of the phase coherence lengths away from the CNP, i.e., with increasing charge carrier density, is a hallmark of DP-type spin scattering with a large SOC strength of $\lambda_\mathrm{sym} = 2.5$ meV. This value is comparable to those obtained in TMD/Gr heterostructures (1-15 meV)~\cite{Garcia2018May,JiangNano2015,AvsarNature2014,WangNature2015,OmarPRB2018,WangPRX2016} and demonstrates the potential of Bi-based TIs for spin control via SOC.

We gratefully acknowledge support by the Helmholtz Nano Facility \cite{Albrecht2017}. This project has received funding from the European Union's Horizon 2020 research and innovation programme under grant agreement No 785219 (Graphene
Flagship), the Virtual Institute for Topological Insulators (J\"ulich-Aachen-W\"urzburg-Shanghai), and by the Deutsche Forschungsgemeinschaft (DFG) through SPP 1666 (BE 2441/8-2). ICN2 is supported by the Severo Ochoa program from Spanish MINECO (Grant No.~SEV-2013-0295) and funded by the CERCA Programme/Generalitat de Catalunya.

\providecommand{\noopsort}[1]{}\providecommand{\singleletter}[1]{#1}%

\widetext
\pagebreak
\begin{center}
	\textbf{\large Supplemental Material: Proximity-Induced Electronic Properties of Graphene-\bsts Heterostructures}
\end{center}
\setcounter{equation}{0}
\setcounter{figure}{0}
\setcounter{table}{0}
\setcounter{page}{1}
\makeatletter
\renewcommand{\theequation}{S\arabic{equation}}
\renewcommand{\thefigure}{S\arabic{figure}}
\renewcommand{\thesection}{S\arabic{section}}
\renewcommand{\thetable}{S\arabic{table}}
\newcommand{\supplabel}{S}
\renewcommand{\thefigure}{\supplabel\arabic{figure}}
\section{\label{sec:level1}Fabrication of heterostructures fully encapsulated in hexagonal boron nitride}
In the first step, hBN flakes were exfoliated on a highly doped silicon substrate covered by a 285 nm SiO$_2$ layer. Using vapor phase deposition technique~\cite{Ockelmann2015Aug}, BSTS flakes were grown on the hBN. In this approach, a standard CVD furnace with three separate heating zones was used. The source BSTS crystal was heated to a temperature of 650$^{\circ}$C and by using a 50 SCCM flow of Ar, the vapor was carried over the substrates in the second zone where the deposition takes place. The temperature of this zone was kept at 325$^{\circ}$C. The process was carried out for 30 min at a pressure of 120 mbar which was adjusted by a PID controller. After the growth, the ultra-thin flake was identified under optical microscope and using the technique described by Wang \textit{et al.}~\cite{WangScience2013}, a single layer of graphene was transferred on top, resulting in two regions of hBN/Gr/BSTS/hBN and hBN/Gr/hBN. Next, an aluminum hard-mask was used to define two Hall bars on top and the structure was etched using reactive ion etching technique with Ar and CHF$_3$ as the process gases. After removing the aluminum hard-mask with a wet chemical etching step, contacts were patterned using electron-beam lithography followed by evaporation of Cr(5 nm)/Au(120 nm).

\section{Transport measurements on  \MakeLowercase{h}BN/BSTS/\MakeLowercase{h}BN structure }
In order to exclude the contribution of BSTS layer in magneto-transport measurements, a fully encapsulated BSTS flake with the same thickness (2 QLs) was  fabricated as a reference sample (see inset of Fig. S1). As shown in Fig.~S1, no localization signal at low magnetic fields and no quantization at high magnetic fields was observed.

\begin{figure}[h!]
\begin{center}
\includegraphics[width=0.85\textwidth]{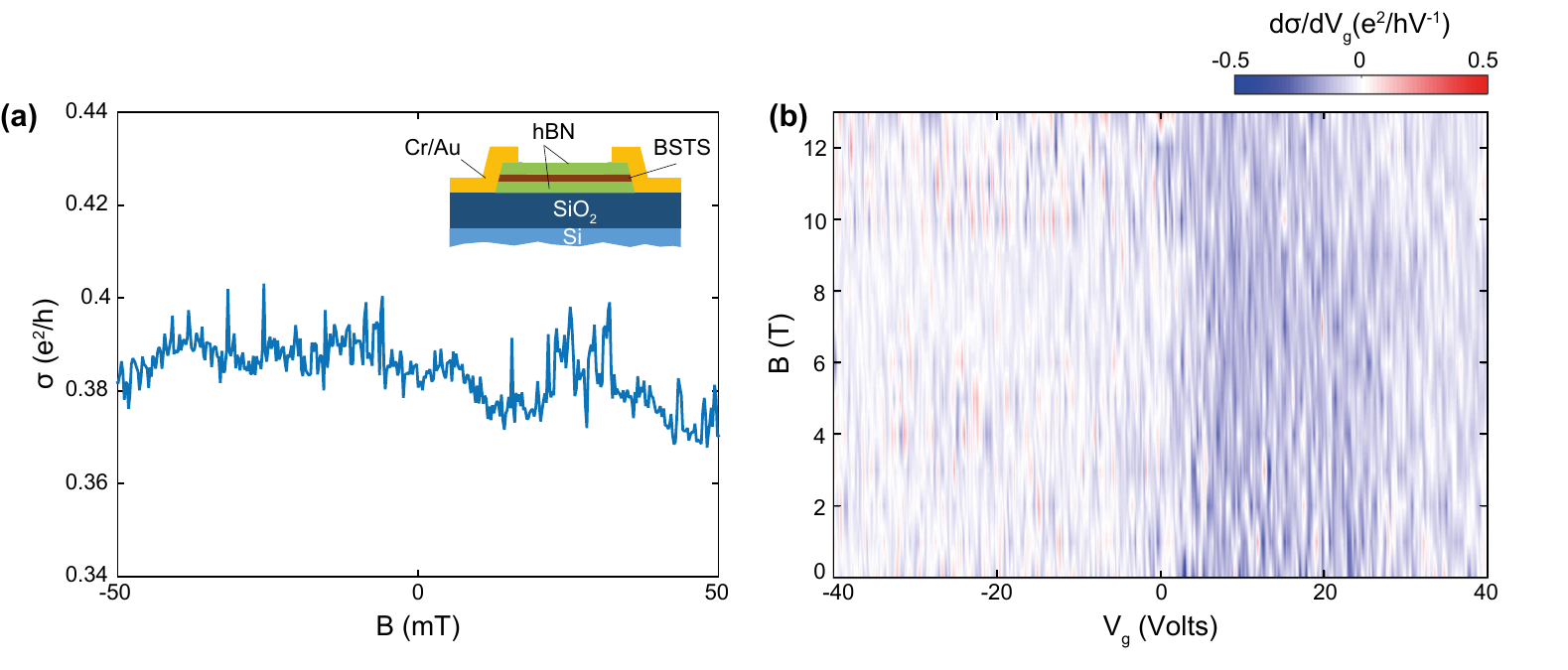}
\caption[figS01]{\label{figS1}
	(a) Low-field magneto-conductivity data of hBN/BSTS/hBN device. (b) Transconductivity of hBN/BSTS/hBN vs gate voltage and perpendicular B field.
}

\end{center}
\end{figure}

\section{Symmetric behavior of G\MakeLowercase{r}/BSTS device with respect to magnetic field}
\begin{figure}[h!]
	\begin{center}
		\includegraphics[width=0.55\textwidth]{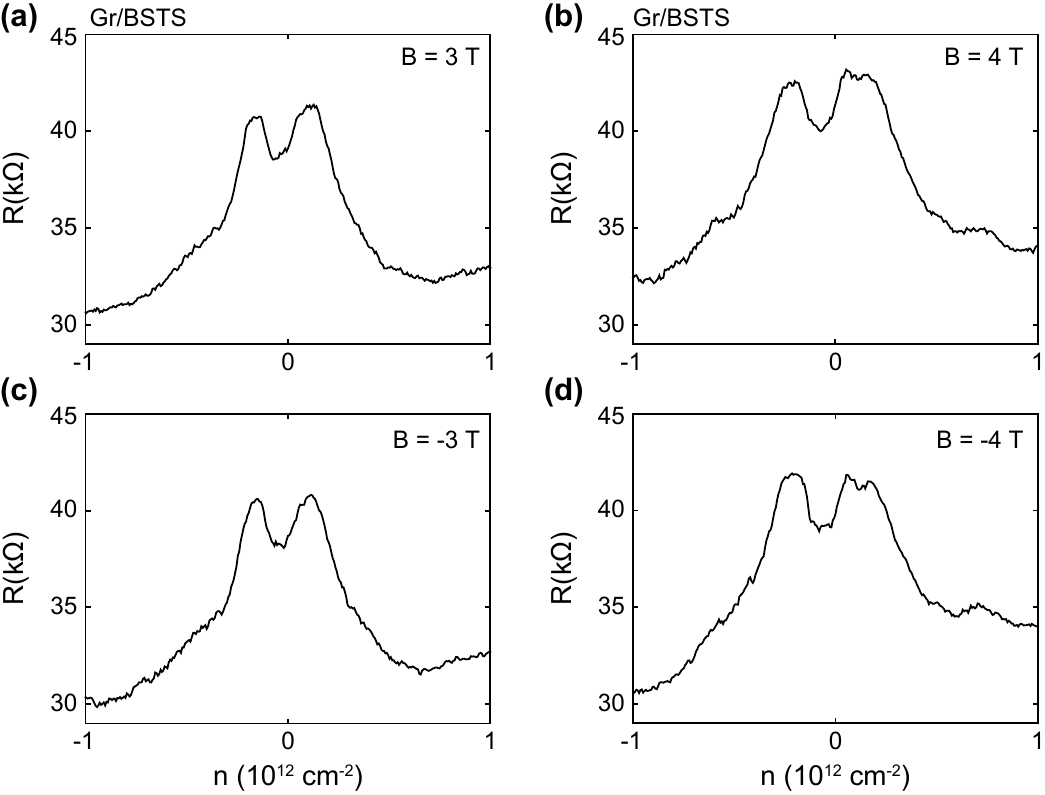}
		\caption[figS01]{\label{figS2}
			Line traces of the Gr/BSTS resistance at (a) 3 T (b) 4 T (c) -3 T (d) -4 T. These measurements were done at a temperature of T=1.7 K.
		}
		
	\end{center}
\end{figure}

The behavior shown in Fig.~2(d) (main text) for magnetic fields of 3 and 4 T can also be observed at negative magnetic fields, in contrast to the observations of Zhang et al.~\cite{ZhangACSNANO2017}, as shown in Fig.~\ref{figS2} for $\pm 3$ and $\pm 4$ T. These measurements were done on the same sample discussed in the main text after being stored in air for several weeks. Therefore, the contacts and most likely the interface of BSTS and Gr have been deteriorated. However, the peculiar behavior at intermediate magnetic fields was still preserved.

\section{WAL in G\MakeLowercase{r}/BSTS device }
\begin{figure}[h!]
	\begin{center}
		\includegraphics[width=0.85\textwidth]{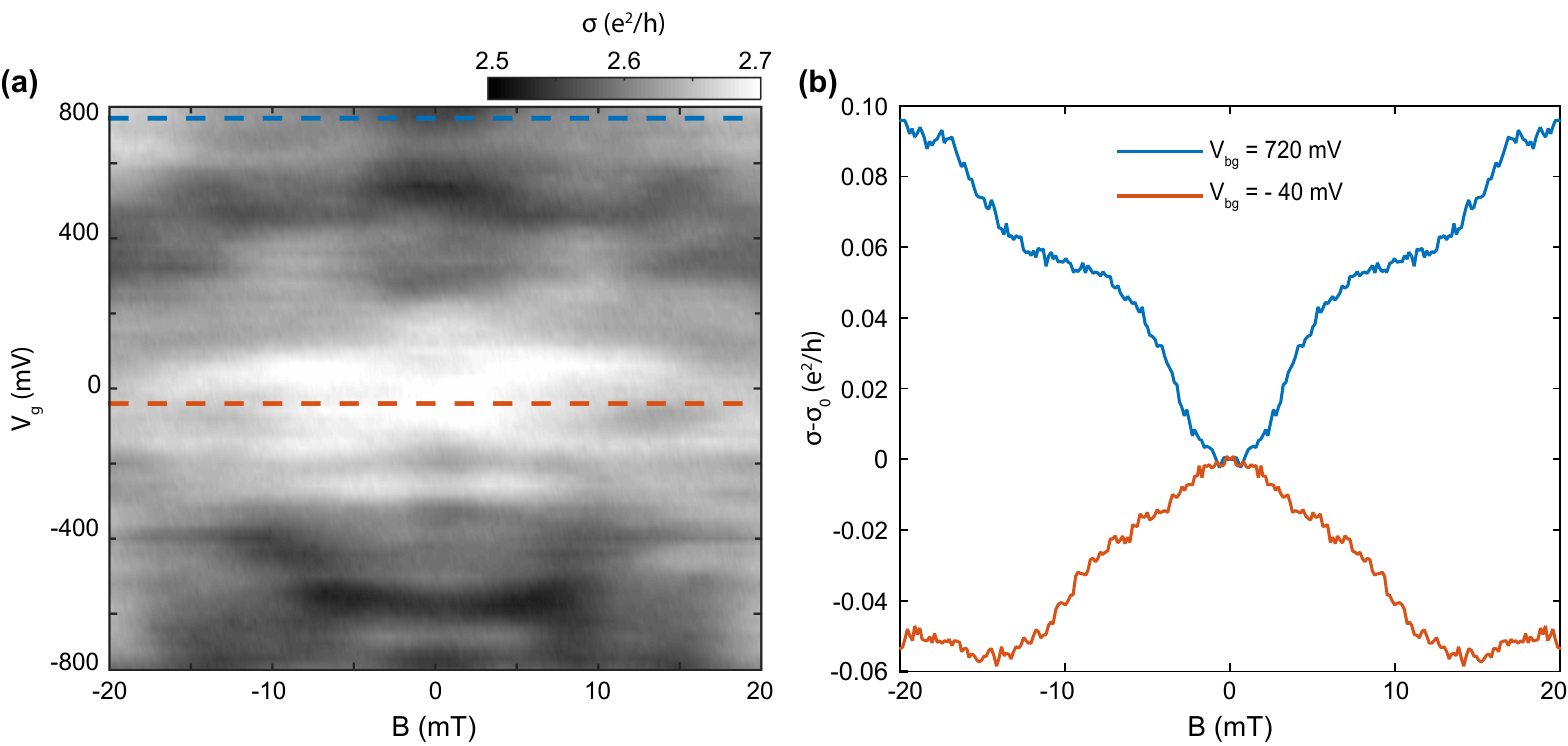}
		\caption[figS01]{\label{figS3}
			(a) Magneto-conductivity of Gr/BSTS as a function of gate voltage showing the transition from WL to WAL. (b) Line traces corresponding to the dashed lines in part (a) which highlight the WL to WAL transition.
		}
		
	\end{center}
\end{figure}

The Gr/BSTS device showed WAL behavior at some specific gate voltages. An example is shown in Fig.~\ref{figS3} where WAL can be seen at a very small range around $V_\mathrm{g}=-40$ mV. This gate voltage corresponds to a charge carrier density of $n = 2.5 \times 10^{11}$ cm$^{-2}$. We have identified several other charge carrier densities including $n = 1.1 \times 10^{12}$ cm$^{-2}$, $n = 2.6 \times 10^{11}$ cm$^{-2}$, $n = 4.2 \times 10^{10}$ cm$^{-2}$ and $n = -6.9 \times 10^{11}$ cm$^{-2}$ where a WAL signal appears. This behavior can be observed up to a temperature of $T\approx 1$ K, above which the WL signal reappears.

\section{Estimation of spin-scattering time scales in G\MakeLowercase{r}/BSTS device }
The complete equation describing W(A)L in graphene can be written as~\cite{McCannPRL2012WL,Garcia2018May}:
\begin{eqnarray}
\Delta \sigma &=&-\frac{{e}^{2}}{2\pi \hbar}\biggl[F\left(\frac{{\tau}_{B}^{-1}}{{\tau}_{\phi}^{-1}}\right)-2F\left(\frac{{\tau}_{B}^{-1}}{{\tau}_{\phi}^{-1}+{\tau}_{\star}^{-1}+{\tau}_{VZ}^{-1}}\right)-F\left(\frac{{\tau}_{B}^{-1}}{{\tau}_{\phi}^{-1}+2{\tau}_{iv}^{-1}}\right)\nonumber\\
&-&2F\left(\frac{{\tau}_{B}^{-1}}{{\tau}_{\phi}^{-1}+{\tau}_{R}^{-1}+{\tau}_{PIA}^{-1}+{\tau}_{I}^{-1}+{\tau}_{VZ}^{-1}}\right)+4F\left(\frac{{\tau}_{B}^{-1}}{{\tau}_{\phi}^{-1}+{\tau}_{*}^{-1}+{\tau}_{R}^{-1}+{\tau}_{PIA}^{-1}+{\tau}_{I}^{-1}}\right)\nonumber\\
&+&2F\left(\frac{{\tau}_{B}^{-1}}{{\tau}_{\phi}^{-1}+2{\tau}_{iv}^{-1}+{\tau}_{R}^{-1}+{\tau}_{PIA}^{-1}+{\tau}_{I}^{-1}+{\tau}_{VZ}^{-1}}\right)-F\left(\frac{{\tau}_{B}^{-1}}{{\tau}_{\phi}^{-1}+2{\tau}_{R}^{-1}+2{\tau}_{PIA}^{-1}}\right)\nonumber\\
&+&2F\left(\frac{{\tau}_{B}^{-1}}{{\tau}_{\phi}^{-1}+{\tau}_{*}^{-1}+2{\tau}_{R}^{-1}+2{\tau}_{PIA}^{-1}+{\tau}_{VZ}^{-1}}\right)+F\left(\frac{{\tau}_{B}^{-1}}{{\tau}_{\phi}^{-1}+2{\tau}_{iv}^{-1}+2{\tau}_{R}^{-1}+2{\tau}_{PIA}^{-1}}\right)\biggl]
\end{eqnarray}

\begin{figure}[h!]
	\begin{center}
		\includegraphics[width=0.85\textwidth]{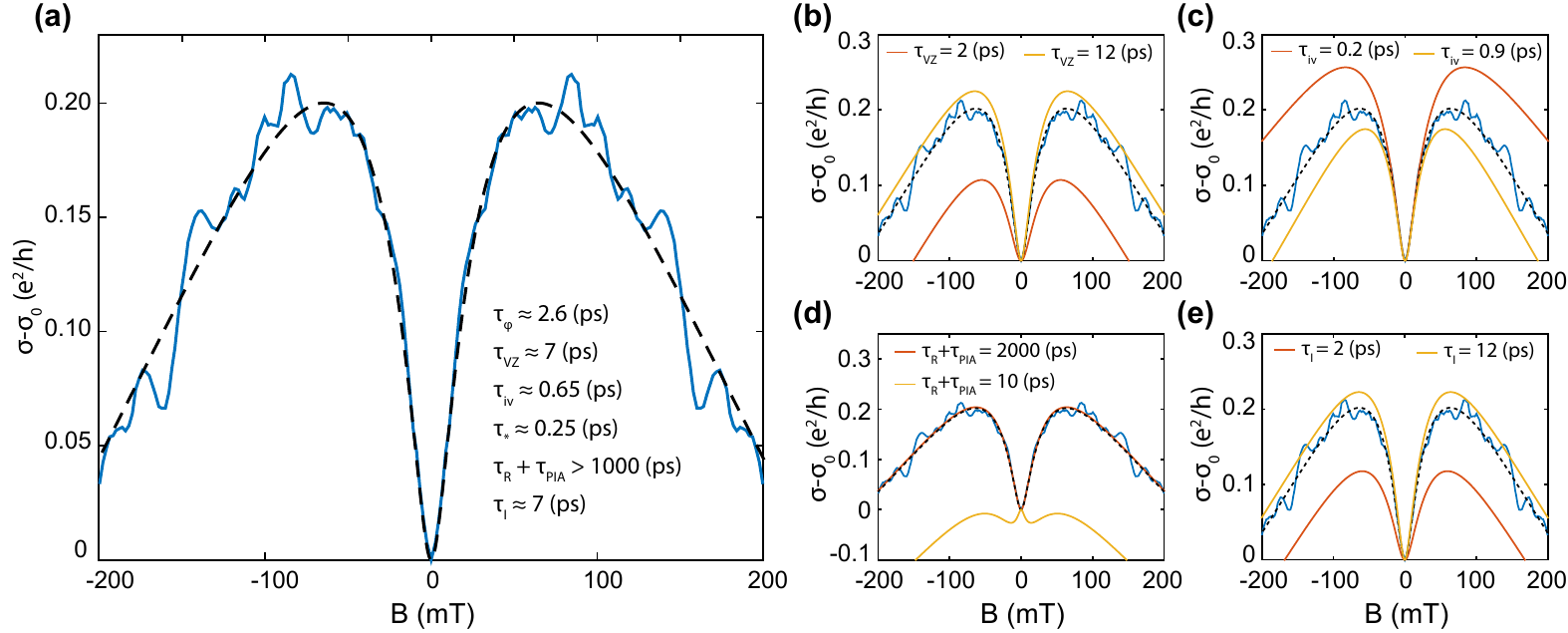}
		\caption[figS01]{\label{figS4}
			(a) Magneto-conductivity of Gr/BSTS device at charge carrier density of $n = 1.2 \times 10^{12}$ cm$^{-2}$. The dashed line shows the calculated values from the extended W(A)L model (Eq.~S1). (b-e) Influence of the (b) valley-Zeeman scattering time (c) inter-valley scattering time (d) asymmetric contributions and (e) intrinsic scattering time on shape of the curve.
		}
		
	\end{center}
\end{figure}

In this model $ \tau_{*}=\tau_\mathrm{iv}+\tau_\mathrm{z}$, where $\tau_\mathrm{iv}$ is the inter-valley and $\tau_\mathrm{z}$ is the intra-valley scattering time. The induced SOC is considered in $\tau_\mathrm{I}, \tau_\mathrm{VZ}, \tau_\mathrm{R}$ and $\tau_\mathrm{PIA}$, which are the intrinsic, valley-Zeeman, Rashba and PIA spin relaxation times respectively. As mentioned in the main text, we do not attempt to fit the WL feature with this model as the large number of fitting parameters results in large uncertainties. However, we can reproduce the measured magneto-conductivity curve by plugging in values for different parameters in order to understand the importance of each spin relaxation mechanism. For this purpose, unlike the fits in the main text, we used a larger magnetic field range since the effect of the additional terms is larger at higher fields. Fig.~\ref{figS4}(a) shows an example trace at a gate voltage of 20~V that can be well described by the model. In ~\ref{figS4}(b-e), the effect of each parameter can be seen. From this analysis we can conclude that the asymmetric SOC ($\tau_\mathrm{R}$ and $\tau_\mathrm{PIA}$) is not playing an important role (Fig.~\ref{figS4}(d)) while the symmetric contributions ($\tau_\mathrm{VZ}$ and $\tau_\mathrm{I}$) are crucial for reproducing shape of the WL curves (Fig.~\ref{figS4}(b and e)).

\begin{thebibliography}{62}
\bibitem [{Sup()}]{SupMat}%
  \BibitemOpen
  \href@noop {} {}\bibinfo {howpublished} {See Supplemetal Material [URL will
  be inserted by publisher] for additional details on the ....}\BibitemShut
  {Stop}%
\end{thebibliography}

\end{document}